\documentclass[aps, prd, superscriptaddress, twocolumn, preprintnumbers,nofootnoteinbib]{revtex4-1}
\usepackage[T1]{fontenc}
\usepackage{microtype}
\usepackage[textsize=scriptsize,backgroundcolor=red!70,linecolor=red]{todonotes}
\usepackage{booktabs}
\usepackage{dcolumn}
\usepackage{amsmath}
\usepackage{amsfonts}
\usepackage{amssymb}
\usepackage{graphicx}
\usepackage{hyperref}
\usepackage[all]{hypcap}
\usepackage{paralist}
\usepackage{multirow}

\usepackage{graphicx}% Include figure files
\usepackage{dcolumn}% Align table columns on decimal point
\usepackage{bm}% bold math
\usepackage{epsf}% Include figure files
\usepackage{dcolumn}% Align table columns on decimal point
\def\be{\begin{equation}}
\def\ee{\end{equation}}
\def\bea{\begin{eqnarray}}
\def\eea{\end{eqnarray}}
\def\gsim{\ \rlap{\raise 2pt\hbox{$>$}}{\lower 2pt \hbox{$\sim$}}\ }
\def\lsim{\ \rlap{\raise 2pt\hbox{$<$}}{\lower 2pt \hbox{$\sim$}}\ }
\def\dslash{\kern-4pt \not{\hbox{\kern-2pt $\partial$}}}
\def\pslash{\not{\hbox{\kern-2pt p}}}

\begin{document}
% \ifpdf
%\DeclareGraphicsExtensions{.pdf,.jpg,.mps,.png}
% \else
\DeclareGraphicsExtensions{.eps,.ps}
% \fi

%\preprint{ROME1-1364-2003}

\title{Degeneracies in long-baseline neutrino experiments from nonstandard interactions}


%%%%%%%%%%%%%%%%%%%%%%%%%%%%%%%%%%%%%%%%%%%%%%%%%%%%%

\author{Jiajun Liao}
%\email[Email Address: ]{liaoj@hawaii.edu}
\affiliation{Department of Physics and Astronomy, University of Hawaii at Manoa, Honolulu, HI 96822, USA}
 
\author{Danny Marfatia}
%\email[Email Address: ]{dmarf8@hawaii.edu}
\affiliation{Department of Physics and Astronomy, University of Hawaii at Manoa, Honolulu, HI 96822, USA}

\author{Kerry Whisnant}
%\email[Email Address: ]{whisnant@iastate.edu}
\affiliation{Department of Physics and Astronomy, Iowa State University, Ames, IA 50011, USA}

%\collaboration can be followed by \email, \homepage, 
%\thanks as well.
%\date{\today}
%%%%%%%%%%%%%%%%%%%% abstract %%%%%%%%%%%%%%%%%%%%%%%%%%%%%%%%%%%%%%%%%%
\begin{abstract}
We study parameter degeneracies that can occur in long-baseline neutrino
appearance experiments due to nonstandard interactions (NSI) in neutrino propagation. 
For a single off-diagonal NSI parameter, and neutrino and antineutrino measurements at a 
single $L/E$, there exists a continuous four-fold degeneracy (related to the mass hierarchy and $\theta_{23}$ octant) that renders the mass hierarchy, octant, and CP phase unknowable. Even with a combination of NO$\nu$A and T2K data, which in principle can resolve the degeneracy, both NSI and the CP phase remain unconstrained because of experimental uncertainties. 
 A wide-band beam experiment like DUNE will resolve this degeneracy if the nonzero off-diagonal
 NSI parameter is $\epsilon_{e\mu}$. If $\epsilon_{e\tau}$ is nonzero, or the diagonal NSI parameter $\epsilon_{ee}$ is ${\cal O}(1)$, a wrong determination of the mass hierarchy and of CP violation can occur at DUNE. The octant degeneracy can be further complicated by $\epsilon_{e\tau}$, but is not 
 affected by $\epsilon_{ee}$.

\end{abstract}
%{11.30.Er.,11.30.Cp.,14.60.Pq,13.15.+g}
%{11.30.Er,11.30.Pb,12.60.Jv}
%\keywords{}
%%%%%%%%%%%%%%%%%%%%%%%%%%%%%%%%%%%%%%%%%%%%%%%%%%%%%%%%%%%%%%%%%%%%%%%%%%
%\maketitle must follow title, authors, abstract, 
%\pacs, and \keywords
\maketitle

%-----------------------------------------------------------%
%%%%%%%%%%%%%%%%%%% SECTION 1 : introduction   %%%%%%%%%%%%%%
%-----------------------------------------------------------%
%\section{Introduction}
%-----------------------------------------------------------%
%%\underline{\bf {Introduction:}} 
After decades of neutrino oscillation experiments, the standard model
(SM) with massive neutrinos is well established and the study of
neutrino oscillations has entered the precision
era~\cite{Agashe:2014kda}. Next generation neutrino oscillation experiments 
will be sensitive to physics beyond the SM, which is often described in a model-independent manner by nonstandard interactions (NSI);
for a recent review see Ref.~\cite{Ohlsson:2012kf}. In general, NSI not only affect neutrino propagation in matter via
neutral-current interactions, but also affect neutrino production and
detection via charged-current interactions. 
Model-independent bounds on the production and detection NSI are
generally an order of magnitude stronger than the matter
NSI~\cite{Biggio:2009nt}. Therefore, we neglect production and
detection NSI in this work, and focus on matter NSI, which
can be described by dimension-six four-fermion operators
of the form~\cite{Wolfenstein:1977ue}
\be
  \label{eq:NSI}
  \mathcal{L}_\text{NSI} =2\sqrt{2}G_F
   \epsilon^{\mathfrak{f}C}_{\alpha\beta} \!
        \left[ \overline{\nu}_\alpha \gamma^{\rho} P_L \nu_\beta \right] \!\!
        \left[ \bar{\mathfrak{f}} \gamma_{\rho} P_C \mathfrak{f} \right] + \text{h.c.}\,,
\ee
where $\alpha, \beta=e, \mu, \tau$, $C=L,R$, $\mathfrak{f}=u,d,e$, and
$\epsilon^{\mathfrak{f}C}_{\alpha\beta}$ are dimensionless parameters that quantify the
strength of the new interaction relative to the SM. Since neutral-current
interactions affect neutrino propagation coherently, the matter
NSI potentially have a large effect on the long-baseline neutrino
oscillation experiments, T2K~\cite{Abe:2011ks}, NO$\nu$A~\cite{Ayres:2004js},
and DUNE~\cite{Adams:2013qkq}. 
Previous studies of matter NSI in these experiments can be found in Refs.~\cite{Coelho:2012bp,coloma}.

In this paper, we use bi-probability plots and numerical simulations to analyze parameter degeneracies in
long-baseline neutrino appearance experiments that arise from matter NSI. We specifically study how well the NO$\nu$A, T2K and DUNE experiments will resolve these degeneracies for the cases of one and two off-diagonal NSI, and diagonal NSI.

{\bf Oscillation probabilities.}
The Hamiltonian for neutrino propagation in the flavor basis may be written as
\be
H = \frac{1}{2E} \left[ U\text{diag}(0,\delta m^2_{21},\delta m^2_{31})
U^\dagger + V\right]\,,
\ee
where $U$ is the PMNS mixing matrix~\cite{Agashe:2014kda}, $\delta m^2_{ij}=m^2_i-m^2_j$,
%\be
%U = \left(\begin{array}{ccc}
%c_{13} c_{12} & c_{13} s_{12} & s_{13} e^{-i\delta}
%\\
%-s_{12} c_{23} - c_{12} s_{23} s_{13} e^{i\delta} &
%c_{12} c_{23} - s_{12} s_{23} s_{13} e^{i\delta} &
%c_{13} s_{23}
%\\
%s_{12} s_{23} - c_{12} c_{23} s_{13} e^{i\delta} &
%-c_{12} s_{23} - s_{12} c_{23} s_{13} e^{i\delta} &
%c_{13} c_{23}
%\end{array} \right)\,,
%\ee
and $V$ represents the potential arising from interactions of neutrinos
in matter,
\be
V = A \left(\begin{array}{ccc}
1 + \epsilon_{ee} & \epsilon_{e\mu}e^{i\phi_{e\mu}} & \epsilon_{e\tau}e^{i\phi_{e\tau}}
\\
\epsilon_{e\mu}e^{-i\phi_{e\mu}} & \epsilon_{\mu\mu} & \epsilon_{\mu\tau}e^{i\phi_{\mu\tau}}
\\
\epsilon_{e\tau}e^{-i\phi_{e\tau}}& \epsilon_{\mu\tau}e^{-i\phi_{\mu\tau}} & \epsilon_{\tau\tau}
\end{array}\right)\,.
\ee
Here,
$A \equiv 2\sqrt2 G_F N_e E$ and $\epsilon_{\alpha\beta}e^{i\phi_{\alpha\beta}}\equiv\sum\limits_{\mathfrak{f},C}\epsilon^{\mathfrak{f}C}_{\alpha\beta}\frac{N_\mathfrak{f}}{N_e}$, with $N_\mathfrak{f}$ the number density of fermion $\mathfrak{f}$. The $\epsilon_{\alpha\beta}$ are real, and $\phi_{\alpha\beta}=0$ for $\alpha=\beta$.
The unit contribution to the $ee$ element of the matrix is due to the standard charged-current
interaction.

Expanding the $\nu_\mu \to \nu_e$ oscillation probability
for the normal hierarchy (NH) in the small quantities $s_{13}$, $r$, and the nondiagonal $\epsilon$, we find (with $c_{jk} \equiv \cos\theta_{jk}$, $s_{jk} \equiv \sin\theta_{jk}$)
\bea
&&P(\nu_\mu \to \nu_e) = x^2 f^2 + 2xyfg \cos(\Delta + \delta) + y^2 g^2
\nonumber\\
&+& 4\hat A \epsilon_{e\mu}
\left\{ xf [s_{23}^2 f \cos(\phi_{e\mu}+\delta)  
+ c_{23}^2 g \cos(\Delta+\delta+\phi_{e\mu})]\right.
\nonumber\\
&\phantom{+}&\qquad\qquad \left.
+yg [c_{23}^2 g \cos\phi_{e\mu} + s_{23}^2 f \cos(\Delta-\phi_{e\mu})]\right\}
\nonumber\\
&+& 4\hat A \epsilon_{e\tau} s_{23} c_{23}
\left\{ xf [f \cos(\phi_{e\tau}+\delta)  
- g \cos(\Delta+\delta+\phi_{e\tau})] \right.
\nonumber\\
&\phantom{+}&\qquad\qquad\qquad \left.
-yg [g \cos\phi_{e\tau} - f \cos(\Delta-\phi_{e\tau})]\right\}
\nonumber\\
&+& 4 \hat A^2 g^2 c_{23}^2 |c_{23} \epsilon_{e\mu} - s_{23}\epsilon_{e\tau}|^2
 +  4 \hat A^2 f^2 s_{23}^2 |s_{23} \epsilon_{e\mu} + c_{23}\epsilon_{e\tau}|^2
\nonumber\\
&+& 8 \hat A^2 fg s_{23} c_{23}
\left\{ c_{23}\cos\Delta
\left[ s_{23}(\epsilon_{e\mu}^2 - \epsilon_{e\tau}^2)\right.\right.
\nonumber\\
&& \qquad\qquad\qquad  \left.\left.+2 c_{23} \epsilon_{e\mu}\epsilon_{e\tau}
\cos(\phi_{e\mu}-\phi_{e\tau})\right]\right.
\nonumber\\
&\phantom{+}& \qquad\qquad\qquad  \left.-\epsilon_{e\mu}\epsilon_{e\tau}
\cos(\Delta-\phi_{e\mu}+\phi_{e\tau})\right\}
\nonumber\\
&+& {\cal O}(s_{13}^2 \epsilon, s_{13}\epsilon^2, \epsilon^3)\,,
\label{eq:prob}
\eea
where following Ref.~\cite{Barger:2001yr},
\bea
x &\equiv& 2 s_{13} s_{23}\,,\quad
y \equiv 2r s_{12} c_{12} c_{23}\,,
\quad r = |\delta m^2_{21}/\delta m^2_{31}|\,,
\nonumber\\
%f &\equiv& {\sin[\Delta(1-\hat A(1+\epsilon_{ee}))]\over(1-\hat A(1+\epsilon_{ee}))}\,,\quad
f,\, \bar{f} &\equiv& \frac{\sin[\Delta(1\mp\hat A(1+\epsilon_{ee}))]}{(1\mp\hat A(1+\epsilon_{ee}))}\,,\ 
g \equiv \frac{\sin(\hat A(1+\epsilon_{ee}) \Delta)}{\hat A(1+\epsilon_{ee})}\,,\nonumber\\
\Delta &\equiv &\left|\frac{\delta m^2_{31} L}{4E}\right|,\ 
\hat A \equiv \left|\frac{A}{\delta m^2_{31}}\right|\,.
\label{eq:define}
\eea
Note that our definitions of $f$, $\bar{f}$, and $g$ here include
$\epsilon_{ee}$, which is not treated as a small parameter. We have set $c_{13} \to 1$, which
is accurate up to first order in $\theta_{13}$. The antineutrino
probability $\overline{P}\equiv P(\overline{\nu}_e \to \overline{\nu}_\mu)$, is given by Eq.~(\ref{eq:prob}) with
$\hat A \to - \hat A$ (and hence $f \to \bar{f}$),
$\delta \to - \delta$, and $\phi_{\alpha\beta} \to -
\phi_{\alpha\beta}$. For the inverted hierarchy (IH), \mbox{$\Delta \to -
\Delta$}, $y \to -y$, $\hat A \to - \hat A$ (i.e., $f
\leftrightarrow - \bar{f}$, and $g \to -g$). Since $s_{13}$ and
$r$ are small, so are $x$ ($\approx 0.2$) and $y$ ($\approx
0.02)$.  Furthermore, $\hat A \lsim 0.3$ for $L \le 1300$~km.  Our
result agrees with the ${\cal O}(\epsilon)$ expression of
Ref.~\cite{Kopp:2007ne}. The 90\% C.L.
limits on the NSI parameters that appear in Eq.~(\ref{eq:prob}) are
$\epsilon_{ee} < 4.2$, $\epsilon_{e\mu} < 0.33$, and $\epsilon_{e\tau}
< 3.0$~\cite{Biggio:2009nt}. The other $\epsilon_{\alpha\beta}$ do not appear in Eq.~(\ref{eq:prob}) up to second order in $\epsilon$. Since Eq.~(\ref{eq:prob}) is only valid for small nondiagonal $\epsilon$, in our simulations the oscillation probabilities are evaluated numerically without approximation.

\begin{figure}
%\captionsetup{singlelinecheck=on}
%\centering
\includegraphics[width=0.45\textwidth]{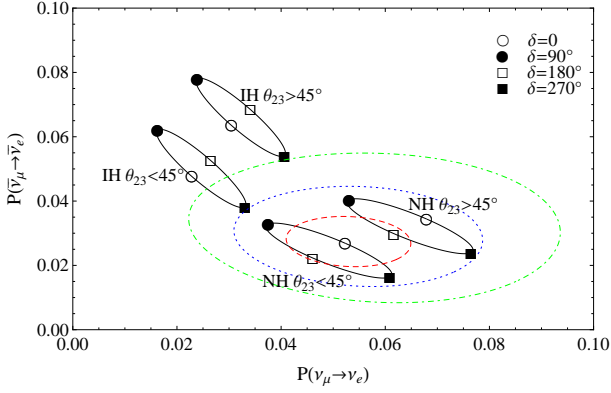}
\caption{Bi-probability plots ($P(\overline{\nu}_\mu\to\overline{\nu}_e)$ versus
$P(\nu_\mu \to \nu_e)$) for $L = 1300$~km and $E=3$~GeV. The solid curves
show the SM values for fixed mixing angles and varying $\delta$ (one curve
for each combination of hierarchy and $\theta_{23}$ octant); the dashed (dotted) [dotdashed]
curves show the values assuming NSI with $\epsilon_{e\mu} =$~0.05 (0.10) [0.15], and varying $\phi_{e\mu}$ for the NH and first octant with
$\delta^\prime = 0$. The mixing angles and mass-squared differences are set
at their best-fit values from Ref.~\cite{Gonzalez-Garcia:2013usa}.}
\label{fig:1}
\end{figure}

%\section{One off-diagonal NSI}
{\bf One off-diagonal NSI.}
If $\epsilon_{ee}=0$, and only one off-diagonal NSI 
 in Eq.~(\ref{eq:prob}) (i.e., $\epsilon_{e\mu}$ or
$\epsilon_{e\tau}$) is nonzero, and a measurement of the neutrino and
antineutrino oscillation probability is made at one particular $L$ and
$E$ (which is approximately true for a narrow-band beam experiment),
then a parameter degeneracy between the SM and a model with NSI will
occur when $P^{SM}(\delta) = P^{NSI}(\delta^\prime,\epsilon,\phi)$ and
$\overline{P}^{SM}(\delta) =
\overline{P}^{NSI}(\delta^\prime,\epsilon,\phi)$, where $\delta'$ is the Dirac CP phase in the model with NSI. Assuming the three
mixing angles and two mass-squared differences are well-measured by
other experiments, for each value of $\delta$ in the SM there are
three unknowns to be determined that give an off-diagonal NSI degeneracy:
$\delta^\prime$, the magnitude $\epsilon$ and the phase $\phi$. With
only two constraining equations, in general there will be a continuum
of solutions that can give a parameter
degeneracy; i.e., for each value of $\delta$, a solution for
$\epsilon$ and $\phi$ will exist for {\it any} value of
$\delta^\prime$.

Therefore, in the context of one off-diagonal NSI, any
$CP$ phase value is allowed with a single
measurement of $P$ and $\overline{P}$. This may be demonstrated via a
bi-probability plot, which shows $\overline{P}$ versus $P$ (see
Fig.~\ref{fig:1}).  The solid ellipses in Fig.~\ref{fig:1} represent
possible $\overline{P}$ and $P$ values for the SM with
fixed mixing angles and varying $\delta$. The four ellipses are for
NH and first $\theta_{23}$ octant, NH and second octant,
IH and first octant, and, IH and second octant, corresponding to the usual
four-fold degeneracy that remains now that the oscillation amplitudes and
mass-squared differences have been measured.

The non-solid contours in Fig.~\ref{fig:1} represent the probabilities for
the same mixing angles for the NH and first octant including NSI with
$\delta^\prime=0$, fixed $\epsilon$ and varying $\phi$. The center of the
non-solid contours is located at the $\delta=0$ point of the
NH-first octant solid ellipse. Clearly any point on any of the solid ellipses
(i.e., any $\delta$, either hierarchy, and either octant) can be
obtained by centering the non-solid contours on any other point of the
NH-first octant solid ellipse (i.e., any $\delta^\prime$) with an
appropriate value of $\epsilon$ and $\phi$. In the limit where the
$\epsilon^2$ terms can be ignored, the non-solid contours are simple
ellipses, the sizes of which are determined by the magnitude
$\epsilon$; when $\epsilon$ is larger, the ellipses become distorted,
their range increasing with $\epsilon$.
NSI with the IH and/or second octant  can be obtained similarly by centering the non-solid contours on the points of the corresponding solid ellipse.

\begin{figure}
\centering
\includegraphics[width=0.45\textwidth]{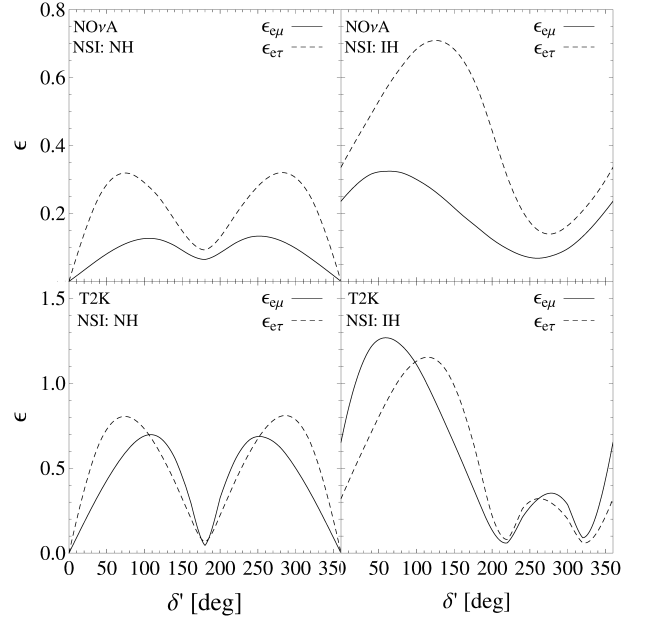}
\caption{Values of a single nonzero $\epsilon$ as a function of $\delta^\prime$ that give $P(\nu_\mu \to \nu_e)$ and
$P(\overline{\nu}_\mu \to \overline{\nu}_e)$ degenerate with the SM
with $\delta = 0$ and NH, at NO$\nu$A ($L=810$~km, $E=2$~GeV) and T2K ($L=295$~km, $E=0.6$~GeV). The mixing angles and mass-squared differences are the best-fit values in Ref.~\cite{Gonzalez-Garcia:2013usa}.
%at (a) $L = 810$~km and $E = 2$~GeV for the NH, (b) $L = 810$~km and $E = 2$~GeV for the IH, (c) $L = 295$~km and $E = 0.6$~GeV for the NH, and (d) $L = 295$~km
%and $E = 0.6$~GeV for the IH. The solid (dashed) curve is for $\epsilon_{e\mu}$
%($\epsilon_{e\tau}$).}
}
\label{fig:2}
\end{figure}

%\subsection{Predictions of degeneracy}
Figure~\ref{fig:2} shows the values of $\epsilon$ versus
$\delta^\prime$ that have a degeneracy with the SM with $\delta = 0$ and NH,
for either $\epsilon_{e\mu}$ or $\epsilon_{e\tau}$, for the baseline
and approximate position of the spectrum peak in the NO$\nu$A and T2K
experiments, assuming NSI solutions with a NH or
IH. For NO$\nu$A, these values are within the corresponding
experimental constraints for all $\delta^\prime$. The values of
$\epsilon$ that give a degeneracy are larger in T2K since the relative
size of the matter effect is smaller there due to the lower average
density and shorter distance. 
%only for $|\delta^\prime| \lsim 40^o$ or
%$|\delta^\prime - 180^o| \lsim 20^o$ for the NH, or $\delta^\prime \gsim 180^o$ for the IH is $\epsilon_{e\mu}$ below its
%constraint.
Similar curves exist for other values of $\delta$ for the NH; in all cases,
$\epsilon = 0$ when $\delta^\prime = \delta$, but $\epsilon > 0$ when
$\delta^\prime \ne \delta$, i.e., degenerate NSI solutions only exist
when $\delta^\prime \ne \delta$. For the IH, the values of $\epsilon$ that give a degeneracy at $\delta^\prime=270^\circ$ are smaller than those at $\delta^\prime=90^\circ$.  This can be understood from the bi-probability plot in Fig.~\ref{fig:1}: the IH with $\delta = 270^\circ$ is closest to the NH. 

Approximate formulas for the curves in the left panels of Fig.~\ref{fig:2}
can be obtained by dropping the $r\epsilon$ and
$\epsilon^2$ terms in Eq.~(\ref{eq:prob}):
\bea
&&\tan(\phi_{e\mu}+\delta^\prime) ={s_{23}^2 (f-\bar{f})\over 2c_{23}^2 g \sin\Delta}
\nonumber\\
& &+{\cos\Delta(\cos\delta-\cos\delta^\prime)[2c_{23}^2 g \cos\Delta
+ s_{23}^2(f+\bar{f})]\over
2c_{23}^2 g\sin^2\Delta(\sin\delta-\sin\delta^\prime)}\,,\label{eq:6}\\
&&\epsilon_{e\mu} = 
%\nonumber\\&&
{yg[\cos(\Delta+\delta)-\cos(\Delta+\delta^\prime)]\over
2\hat A[c_{23}^2 g \cos(\Delta+\phi_{e\mu}+\delta^\prime)
+ s_{23}^2 f\cos(\phi_{e\mu}+\delta^\prime)]}\,,\nonumber
\eea
 and
\bea
&&\tan(\phi_{e\tau}+\delta^\prime) ={(\bar{f}-f)\over 2g \sin\Delta}
\nonumber\\
& &+{\cos\Delta(\cos\delta-\cos\delta^\prime)[2 g \cos\Delta - f -\bar{f}]\over
2g\sin^2\Delta(\sin\delta-\sin\delta^\prime)}\,,\label{eq:7}\\
&&\epsilon_{e\tau} = 
%\nonumber\\&&
{-yg[\cos(\Delta+\delta)-\cos(\Delta+\delta^\prime)]\over
2\hat A s_{23} c_{23} [g \cos(\Delta+\phi_{e\tau}+\delta^\prime)
- f\cos(\phi_{e\tau}+\delta^\prime)]}\,.\nonumber
\eea
$\epsilon_{e\mu}$ and  $\epsilon_{e\tau}$ are obtained by first solving Eqs.~(\ref{eq:6}) and~(\ref{eq:7}) for $\phi_{e\mu}$ and $\phi_{e\tau}$.
Similar equations exist for each of the other
possibilities, i.e., NSI with any $\delta^\prime$, either hierarchy and
either octant can mimic the SM with any $\delta$, either hierarchy
and either octant.

To remove the NSI degeneracies, additional measurements must be made.
Since there are degeneracies throughout the two-dimensional
$\delta-\delta^\prime$ space, one additional measurement (such as a
neutrino probability at a different $L$ and/or $E$) will reduce the
dimensionality of the degeneracies by one, i.e., to lines in
$\delta-\delta^\prime$ space. Thus for each value of $\delta$ there
will only be one $\delta^\prime$ that will be degenerate (or perhaps a
finite number of $\delta^\prime$ if there are multiple solutions).
Two additional measurements at a different $L$ and/or $E$ will remove
the degeneracies; the only solutions then are $\delta^\prime = \delta$ and
$\epsilon = 0$.

\begin{figure}
\centering
\includegraphics[width=0.45\textwidth]{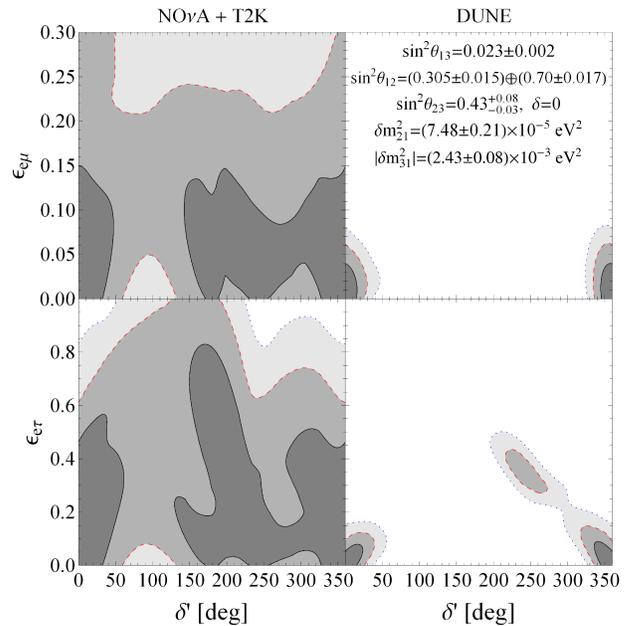}
\caption{$1\sigma$, $2\sigma$ and $3\sigma$ allowed regions for 
$\epsilon_{e\mu}$ and $\epsilon_{e\tau}$ (when only one of them is nonzero) from combined NO$\nu$A and T2K data, and from DUNE.
%the NSI parameter (a) $\epsilon_{e\mu}$ versus $\delta^\prime$ in the combined NO$\nu$A and T2K, (b) $\epsilon_{e\mu}$ versus $\delta^\prime$ in the DUNE, (c) $\epsilon_{e\tau}$ versus $\delta^\prime$ in the combined NO$\nu$A and T2K, and (d)  $\epsilon_{e\tau}$ versus $\delta^\prime$ in the DUNE experiments if the
%data are consistent with the SM with $\delta=0$ in the NH. The solid (dashed) [dotted]
%curves represent the $1\sigma$ ($2\sigma$) [$3\sigma$] bounds.
}
\label{fig:emoret}
\end{figure}

%\subsection{Simulations}
{\bf Simulations.}
In practice, even narrow-band beams have a spread of energies,
although the energy resolution and uncertainties may not allow the
degeneracies to be resolved. We simulate the long-baseline experiments using the GLoBES software
\cite{GLOBES}, supplemented with the new physics tools developed in
Refs.~\cite{Kopp:2007ne, Kopp:2006wp}. 
%The oscillation probabilities are evaluated numerically without approximation. 
We use the experimental setup for NO$\nu$A and T2K as in Ref.~\cite{Huber:2009cw}, in which T2K collects data for 5 years each in the neutrino and antineutrino mode, while NO$\nu$A runs for 3 years in each mode. For DUNE, we consider a 34 kiloton liquid argon detector with a 1.2 MW beam, and running for 3 years in each mode. We checked that our measurement precision is comparable to the projected results in Ref.~\cite{Adams:2013qkq}, and the expected sensitivity for the mixing angles and mass-squared differences in our simulation match Fig.~8 of Ref.~\cite{Berryman:2015nua}. The Preliminary Reference Earth Model density profile~\cite{Dziewonski:1981xy} is implemented in GLoBES, and we use a conservative 5\% uncertainty for the matter density~\cite{Geller:2001ix}. Also, to be conservative, we adopt the central values and priors on the mixing angles and mass-squared differences, and the sizes of the NSI parameters in our simulation from the
global-fit with NSI in Ref.~\cite{Gonzalez-Garcia:2013usa}. 

\begin{figure}
\centering
\includegraphics[width=1.6in]{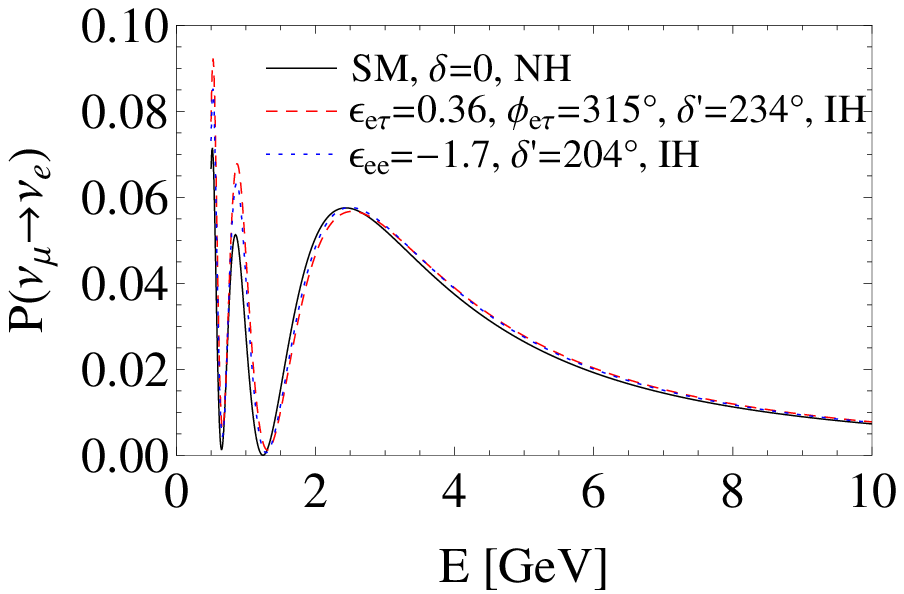}
\includegraphics[width=1.6in]{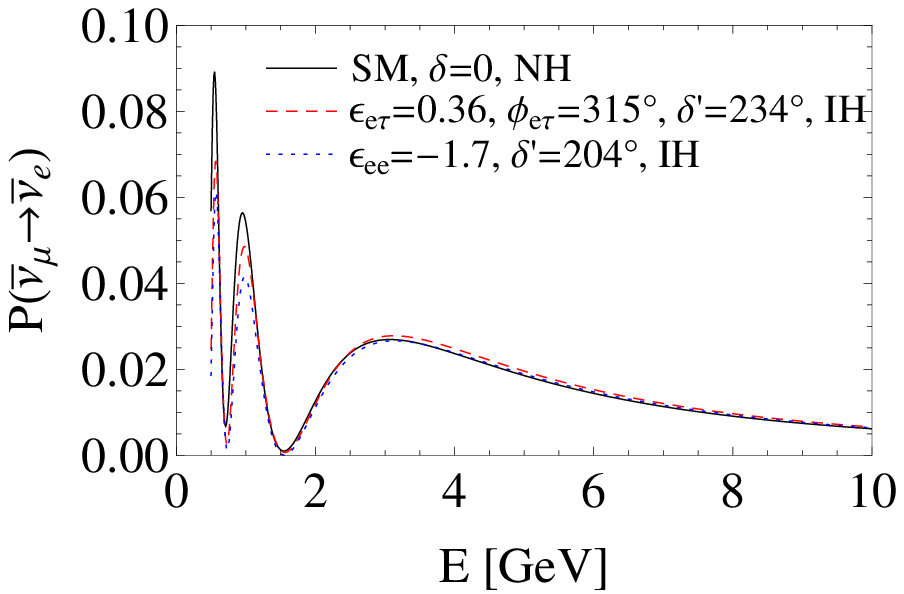}
\caption{$\nu_\mu \rightarrow \nu_e$ and $\overline{\nu}_\mu \rightarrow \overline{\nu}_e$ oscillation probabilities at DUNE. The mixing angles and mass-squared differences are the best-fit values in Ref.~\cite{Gonzalez-Garcia:2013usa}.
}
\label{fig:PvE}
\end{figure}

In Fig.~\ref{fig:emoret} we show the
expected allowed regions (defined by $\Delta \chi^2$ for two degrees of freedom assuming the parameters are Gaussian distributed) in the $\delta^\prime-\epsilon_{e\mu}$ space and $\delta^\prime-\epsilon_{e\tau}$ space from the neutrino and antineutrino appearance
channels from NO$\nu$A and T2K combined, and from DUNE. We assume that the data are consistent with
the SM with $\delta = 0$, the first octant, and the NH. We scan over all octant and hierarchy combinations.
%From Fig.~\ref{fig:3} (a), we see that for all values of $\delta^\prime$ there exist
%NSI solutions with nonzero $\epsilon_{e\mu}$ that are within $2\sigma$ region for NO$\nu$A, 
%consistent with there being a possible NSI solution for any
%value of $\delta^\prime$. In fact, the region allowed at less than
%$1\sigma$ (between the solid curves) tracks well with the solid curve
%in Fig.~\ref{fig:2} after taking both the NH and IH into account. $\delta^\prime = 90^o$ are excluded at less than the $2\sigma$ level for
%$\epsilon_{e\mu}=0$, and $\delta^\prime = 270^o$ are still allowed at less than $1\sigma$ level because the IH requires smaller values of $\epsilon_{e\mu}$ for the degenerate solutions. This can be understood from the
%bi-probability plot in Fig.~\ref{fig:1}: the IH with $\delta = 270^o$ is closest to the NH. 
%A similar plot for the T2K experiment is shown in
%Fig.~\ref{fig:3} (b), and contours for $\epsilon_{e\tau}$ are shown in
%Figs.~\ref{fig:4} (a) and \ref{fig:4} (b). 
%Neither NO$\nu$A or T2K experiment alone can
%conclusively exclude NSI solutions with a single $\epsilon$. 
%By using neutrino and antineutrino measurements from the combined 
Even with the NO$\nu$A and
T2K data combined, there are regions
near $\delta^\prime = 0$ or $180^\circ$ where NSI solutions are allowed at
less than $1\sigma$; see the left panels of Fig.~\ref{fig:emoret}. So, although theoretically the
degeneracies with NSI solutions should be resolved, the experimental
uncertainties are large enough that large NSI regions are not
excluded. Using the wide-band beam at DUNE, which
effectively measures probabilities at a variety of energies, puts
severe restrictions on NSI: $|\delta^\prime| \lsim 50^\circ$ and
$\epsilon \lsim 0.1$ at $3\sigma$ for $\epsilon_{e\mu}$; see the top-right panel of Fig.~\ref{fig:emoret}. 
However, some small degenerate regions in $\delta^\prime-\epsilon_{e\tau}$ space are allowed at less than $2\sigma$ due to the mass hierarchy degeneracy; see the bottom-right panel of Fig.~\ref{fig:emoret}. This can be understood from the oscillation probabilities shown in Fig.~\ref{fig:PvE}. The dashed curves almost overlap the SM curves in both the neutrino and antineutrino channels. 
This could lead to a wrong determination of the mass hierarchy and CP phase.
Also, in parts of the allowed parameter-space, $\theta_{23}$ lies in the second octant which could lead to a wrong determination of the octant.
We see that DUNE alone cannot completely resolve these degeneracies for nonzero $\epsilon_{e\tau}$.

\begin{figure}
\centering
\includegraphics[width=0.45\textwidth]{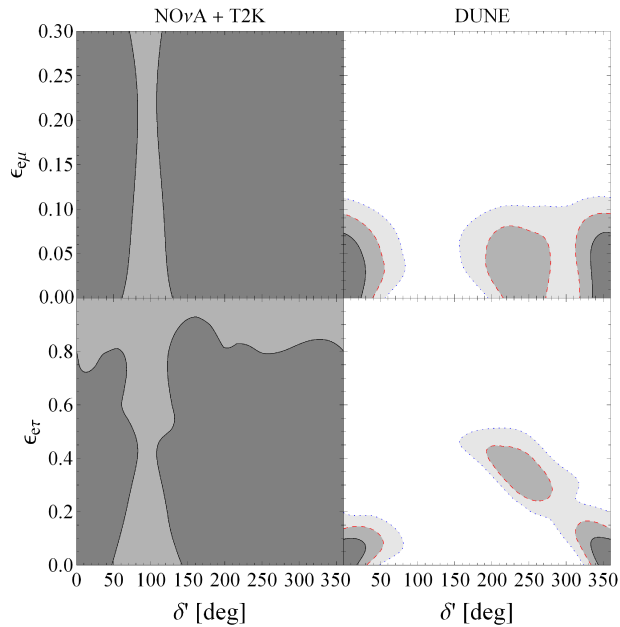}
\caption{Same as Fig.~\ref{fig:emoret}, except both $\epsilon_{e\mu}$ and $\epsilon_{e\tau}$ are nonzero.}
\label{fig:emandet}
\end{figure}

%\section{Two off-diagonal NSI}
{\bf Two off-diagonal NSI.}
If both $\epsilon_{e\mu}$ and $\epsilon_{e\tau}$ are nonzero, then
there are five free NSI parameters: $\delta^\prime$, two magnitudes,
and two phases. Therefore, two $\epsilon$'s and $P$ and $\overline{P}$ measurements at two
different $L$ and $E$ combinations (four equations and five unknowns)
will have potential NSI degeneracies for any $\delta$ and
$\delta^\prime$ (just as with one $\epsilon$ and $P$ and
$\overline{P}$ measurements at just one $L/E$).  The corresponding allowed regions
of $\epsilon_{e\mu}$ and $\epsilon_{e\tau}$ from combining the
NO$\nu$A and T2K experiments are shown in Fig.~\ref{fig:emandet}; 
any value of $\delta^\prime$ is allowed at
less than $2\sigma$.  Therefore $P$ and $\overline{P}$ measurements at
a {\it third} $L/E$ are in principle required to resolve the
degeneracies. Alternatively, a wide-band beam experiment can be used.
Figure~\ref{fig:emandet} shows expected allowed regions
from DUNE. As expected, DUNE cannot resolve all the degeneracies.

\begin{figure}[]
\centering
\includegraphics[width=0.45\textwidth]{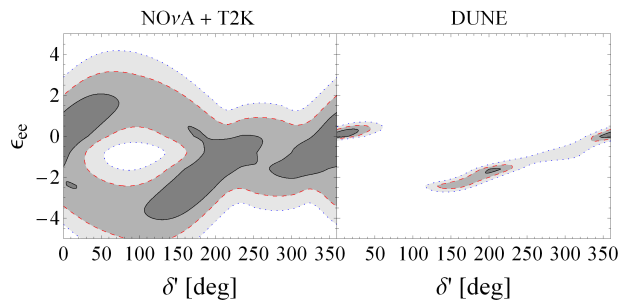}
\caption{Same as Fig.~\ref{fig:emoret}, except for $\epsilon_{ee}$.}
\label{fig:ee}
\end{figure}

%\section{One diagonal NSI ($\epsilon_{ee}$)}
{\bf One diagonal NSI ($\epsilon_{ee}$).}
In this case, the oscillation probability to second order is simply
the first line of Eq.~(\ref{eq:prob}). Since $\phi_{ee}=0$,
there is a simple two-fold degeneracy between the SM and NSI, i.e.,
for measurement of $P$ and $\overline{P}$ at one $L/E$, each value of
$\delta$ is degenerate with a point ($\delta^\prime, \epsilon_{ee}$)
in NSI space (although in some cases, due to the nonlinear nature of
the equations, there are multiple, but finite number of degenerate
solutions). 
%The values of $\epsilon_{ee}$ and $\delta^\prime$ that give
%degenerate solutions are shown versus $\delta$ in Fig.~\ref{fig:Deg-ee}.
In principle, a narrow-band beam experiment should be able to pinpoint
both the SM value of $\delta$ and the degenerate NSI values of
$\delta^\prime$ and $\epsilon_{ee}$. In practice, for NO$\nu$A and
T2K, the uncertainties are too large and no value of $\delta^\prime$
is strongly preferred; see Fig.~\ref{fig:ee}. DUNE will put stronger
constraints on NSI and $\delta^\prime$. The islands in the right panel of Fig.~\ref{fig:ee} can be understood from the dotted curves in Fig.~\ref{fig:PvE}. DUNE alone cannot resolve the mass hierarchy degeneracy, and it could also lead to a wrong determination of CP violation if $\epsilon_{ee}$ is ${\cal O}(1)$. Note that $\epsilon_{ee}$ does not affect the octant degeneracy.

%\section{Summary and conclusions}

In summary, we studied parameter degeneracies that occur in
long-baseline neutrino appearance experiments due to matter NSI. We derived the oscillation probabilities for
the appearance channels to second order in $\epsilon$. We found that there is a continuous four-fold degeneracy for an
off-diagonal NSI in narrow-band beam experiments like NO$\nu$A and T2K. A combination of their data would in principle break the degeneracy, but in practice, large regions of NSI parameter space remain allowed due to large experimental uncertainties. We also discussed degeneracies that occur for diagonal NSI, and for more than one off-diagonal NSI at a time. 
While the DUNE experiment can resolve most of the degeneracies, for nonzero $\epsilon_{e\tau}$ or $\epsilon_{ee}$, there are some parameter regions in which DUNE could lead to a wrong determination of the mass hierarchy and of $CP$ violation. Additionally, for nonzero $\epsilon_{e\tau}$ an incorrect conclusion about the octant of $\theta_{23}$ may be drawn. Nonzero $\epsilon_{ee}$ does not impact a resolution of the octant degeneracy. We conclude that DUNE alone cannot resolve all the degeneracies arising from NSI. (We did not consider the possibility of diagonal and off-diagonal NSI parameters being nonzero simultaneously, which leads to degeneracies between NSI parameters~\cite{coloma}. Clearly, this will further hinder the interpretation of DUNE data.) In this work we focused on how NSI may mimic the SM with $CP$ conservation. In future work we consider degeneracies as a function of the $CP$ phase~\cite{future}. 

% \vspace{0.1 in}
{\it Acknowledgments.} KW thanks the University of Hawaii at Manoa for its hospitality in the
initial stages of this work. This research was supported by the
U.S. DOE under Grant No. DE-SC0010504.

\vskip1cm

%\newpage
%%%%%%%%%%%%%%%%%%%%%%%%%%%%%%%%%%%%%%%%%%%%%%%%%%%%%%%%%%%%

\end{document}